\documentclass[preprintnumbers,showpacs,amsmath,amssymb,floatfix,9pt,prd,twocolumn,
superscriptaddress,nofootinbib]{revtex4}
\usepackage{graphicx}
\usepackage{latexsym}
\usepackage{epsfig}
\usepackage{amssymb}
\usepackage{color}
\usepackage{amsmath}
\usepackage{float}
\usepackage{hyperref}
\begin{document}
\title{Compact star in Tolman-Kuchowicz spacetime in background of Einstein-Gauss-Bonnet gravity }

\author{Piyali Bhar}
\email{piyalibhar90@gmail.com
 } \affiliation{Department of
Mathematics,Government General Degree College, Singur, Hooghly, West Bengal 712 409,
India}

\author{Ksh. Newton Singh}
\email{ntnphy@gmail.com}
\affiliation{Department  of  Physics,  National  Defence  Academy, Khadakwasla,  Pune-411023,  India}

\author{Francisco Tello-Ortiz}
\email{francisco.tello@ua.cl
 } \affiliation{Departamento de F$\acute{i}$sica, Facultad de ciencias b$\acute{a}$sicas,
Universidad de Antofagasta, Casilla 170, Antofagasta, Chile.}

\begin{abstract}\noindent
The present work is devoted to the study of anisotropic compact matter distributions within the framework of 5-dimensional Einstein-Gauss-Bonnet gravity. To solve the field equations, we have considered that the inner geometry is described by Tolman-Kuchowicz spacetime. The Gauss-Bonnet Lagrangian $\mathcal{L}_{GB}$ is coupled to Einstein-Hilbert action through a coupling constant, namely $\alpha$. When this coupling tends to zero general relativity results are recovered. We analyze the effect of this parameter on the principal salient features of the model, such as energy density, radial and tangential pressure and anisotropy factor. These effects are contrasted with the corresponding general relativity results. Besides, we have checked the incidence on important mechanism: equilibrium by means of generalized Tolman-Oppenheimer-Volkoff equation and stability through relativistic adiabatic index and Abreu's criterion. Additionally, the behaviour of the subliminal sound speed of the pressure waves in the principal direction of the configuration and the conduct of the energy-momentum tensor throughout the star are analyzed employing causality condition and energy conditions, respectively. All these subjects are supported by mean of physical, mathematical and graphical survey. The $M-I$ and $M-R$ graphs implies that the stiffness of the equation of state (EoS) increases with $\alpha$, however less stiff than GR.
\end{abstract}
\maketitle

\section{Introduction}
Nowadays it is of great interest to obtain models that describe compact structures, that is, massive objects with a small size, which due to their high density these configurations such as white dwarfs, neutron stars or more exotic stars such as those formed by quarks, constitute a real laboratory to investigate in the regime of strongly coupled gravitational fields. For a long time the development of these models in order to describe and understand the behavior of the aforementioned objects was under the framework of the general relativity theory (GR). With great observational and experimental support \cite{r1} GR describe very well the gravitational interaction and its consequences in a four dimensional spacetime. However, two questions arise: is it possible to study gravity in dimensions smaller than 4-dimensions? and is it possible to study gravity in dimensions greater than 4-dimensions? If so, what benefits and consequences would such studies bring? In the first case, for a 2-dimensional spacetime the Einstein tensor is zero. This is just the consequence of the Einstein-Hilbert Lagrangian being the 2-dimensional Euler characteristic $\chi$. Which is a topological invariant in two dimensions, and therefore we can not obtain equations of motion for our fields from it. In 3-dimensions we can already have an Einstein tensor not identically zero, but we run into another problem. Now the number of independent components of the Riemann tensor is six: the same number of independent components of the Ricci tensor. So, Ricci-plane solutions, i.e., those with $R_{\mu\nu}=0$, are solutions with vanishing Riemann tensor, not giving place to solutions of gravitational waves for example. We then end in the usual three spatial dimensions plus a temporal one. The previous discussion may already be enough to hope that the study in larger dimensions can bear fruit. Perhaps the dynamics resulting from the action of Einstein-Hilbert in four dimensions hide effects that in larger dimensions they could manifest. In fact, several theories have been favored in part to study in larger dimensions, such as the Kaluza-Klein theory (adding an extra dimension to unify gravity with electromagnetism) or string theory (reaching a total of eleven dimensions in in order to unify all the known interactions). Thus, in larger dimensions there is no reason to exclude quadratic, cubic terms, etc. of scalars formed from the Riemann tensor and its contractions. In this direction Lanczos \cite{r2} was the first to extend the GR including covariant high-order derivatives terms of the metric tensor, in order to study the scale invariance under $g_{\mu\nu}\rightarrow \lambda g_{\mu\nu}$ transformation, being $\lambda$ a constant parameter. Nevertheless,the quadratic terms combination found by Lanczos in 4-dimensions did not contribute to the dynamics of the theory. This was because Lanczos was dealing with the four-dimensional Euler characteristic $\chi$, which is a topological invariant in 4-dimensions just as Einstein-Hilbert action is in 2-dimensions.

\begin{figure}[H]
    \centering
        \includegraphics[scale=.6]{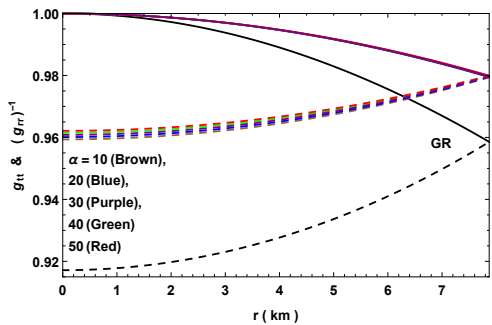}
       \caption{Variation of metric potentials for 4U 1538-52 using the parameter provided in Table I.}
    \label{fig1}
\end{figure}

Lovelock \cite{r3}, later generalized the Einstein-Hilbert action including terms of higher order, with the first-order term corresponding to the Einstein-Hilbert action and the second-order one to the Gauss-Bonnet (GB) Lagrangian. In a n-dimensional spacetime (with $n\geq5$) GB Lagrangian leads to second order equations of motion, as is required. In the spirit of searching for compact structures,  Einstein-Gauss-Bonnet (EGB) theory is promising. In the context of black holes, Boulware and Deser \cite{r4} generalized the higher dimensional solutions in Einstein theory due to Tangherlini \cite{r5}, obtained the exterior vacuum spacetime, \i.e the equivalent Schwarzschild solition in EGB theory. Moreover, the study by Ghosh and Deshkar \cite{r6} about Vaidya radiating black holes in EGB gravity revealed that the location of the horizons is changed as compared to the standard 4-dimensional gravity. In the cosmological and modified gravity theories context EGB gravity has received great attraction \cite{r7,r8,r9,r10,r11,r12,r13,r14,r15,r16,r17}. Recently, Bamba et.al have investigated the energy conditions in the cosmological scenario employing FLRW spacetime. On the other hand, regarding stellar interiors many interesting works available in the literature have been devoted to the study of the existence of collapsed structures \cite{r19,r20,r21,r22}. Besides, Wright \cite{r24} has studied the maximum mass-radius ratio (Buchdahl's limit \cite{r25}) in 5-dimensional EGB gravity.\\

The study of compact object driven by an anisotropic matter distributions, has a long history. Since pioneering work by Bowers and Liang \cite{r26} many researchers have been studying the properties and consequences of this type of structures \cite{r27,r28,r29,r30,r31,r32,r33,r34,r35,r36,r37,r38,r39,r40,r41,r42}. These well-known works explore diverse properties such as: mechanisms of stability and hydrodynamic equilibrium, the behavior of the material content through energy conditions, causality conditions, maximum limit of the mass-radius ratio, maximum value of the superficial redshift, etc. A recent work on the role played by the anisotropy on the properties mentioned above is \cite{r43} (and references contained therein).\\

Following this line, in this paper we construct a well behaved anisotropic fluid sphere in the 5-dimensional EGB scenario, by using Tolman-Kuchowicz \cite{tolman,kucho} spacetime. This metric has been used by other authors in the study of anisotropic charged/uncharged interior solutions \cite{jasim,tello}. So, the plan of this paper is as follows: : In Sec. \ref{sec2} we have discussed about the Einstein Gauss Bonnet gravity in a 5-dimensional spacetime.\\

\begin{figure}[t]
    \centering
        \includegraphics[scale=.6]{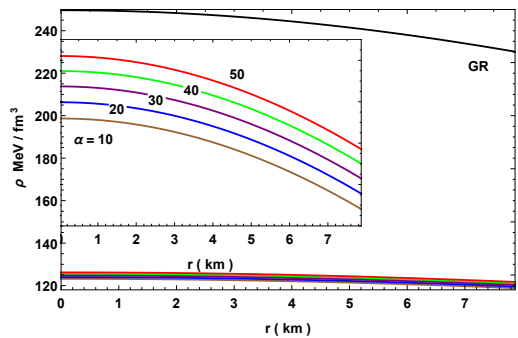}
       \caption{Variation of density for 4U 1538-52 using the parameter provided in Table I.}
    \label{fig2}
\end{figure}

In Sec. \ref{sec3} we have studied the field equations and the mathematical solutions of EGB gravity within Tolman-Kuchowicz spacetime, obtaining the main salient features that characterize the model such as the energy matter density $\rho$, the radial $p_{r}$ and tangential $p_{t}$ pressure and the anisotropy factor $\Delta$. In and Sec. \ref{sec4} we analyze the physical and mathematical behavior of the thermodynamic variables. In Sec. \ref{sec5} we obtain the complete set of constant parameters, joining the inner geometry with exterior spacetime in a smoothly way. Several physical properties have been studied in Secs. \ref{sec6}, \ref{sec7}, \ref{sec8} and \ref{sec9}, such as causality condition, stability, equilibrium under different forces ans¿d energy conditions. Finally in the last Sec. \ref{sec10} we have provided some remarks of the obtained model.

\section{Field Equations}\label{sec2}
We start with a brief description of Einstein-Gauss-Bonnet gravity without cosmological constant \cite{maeda}. The action in the n($\geq5$)-dimensional spacetime is given by
\begin{equation}\label{eq1}
S=\int d^{n}x\sqrt{-g}\left[\frac{1}{2\kappa^{2}_{n}}\bigg(R+\alpha\mathcal{L}_{GB}\bigg)\right] +S_{matter},
\end{equation}
being $\kappa_{n}=\sqrt{8\pi G_{n}}$ and $R$ the n-dimensional Ricci scalar. The last term in Eq. (\ref{eq1}) represents the action for matter fields. The Gauss-Bonnet term (also known as Lovelock's second order term \cite{r3}) compromises the combination of the Ricci scalar R, Ricci tensor $R_{\mu\nu}$, and Riemann tensor $R^{\omega}_{\ \beta\mu\nu}$. Explicity the Gauss-Bonnet terms reads
\begin{equation}\label{eq2}
\mathcal{L}_{GB}=R^{2}-4R_{\mu\nu}R^{\mu\nu}+R_{\omega\beta\mu\nu}R^{\omega\beta\mu\nu}.
\end{equation}
It is worth mentioning that in a four dimensional spacetime the Gauss-Bonnet term (\ref{eq2}) does not contribute to the field equations since it becomes a
total derivative (It is related with a topological invariant, specifically Euler characteristic). It should be noted that when the coupling constant $\alpha$ is zero then GR results are recovered.

\begin{figure}[t]
    \centering
        \includegraphics[scale=.55]{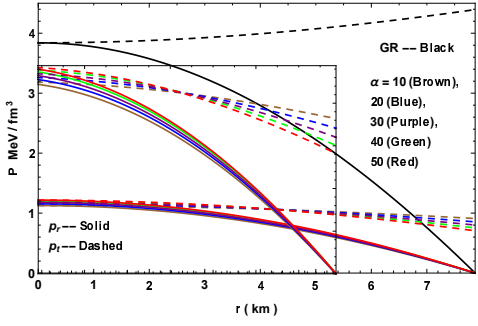}
       \caption{Variation of pressures for 4U 1538-52 using the parameter provided in Table I.}
    \label{fig3}
\end{figure}

The action given by (\ref{eq1}) can be obtained from the lower energy limit in the heterotic string theory. In such case the coupling parameter $\alpha$ is related with the inverse string tension and is positive definite. Therefore, we will consider $\alpha\geq 0$ throughout the study. Variations of (\ref{eq1}) respect to the metric tensor $g_{\mu\nu}$ yields to the following field equations
\begin{equation}\label{eq3}
G_{\mu\nu}+\alpha H_{\mu\nu}=\kappa^{2}_{n}T_{\mu\nu},
\end{equation}
where $G_{\mu\nu}$ and $H_{\mu\nu}$  stand for the Einstein tensor and the Lanczos tensor, respectively. The corresponding expression for these tensor are given by
\begin{equation}\label{eq4}
G_{\mu\nu}=R_{\mu\nu}-\frac{R}{2}g_{\mu\nu},
\end{equation}
\begin{equation}\label{eq5}
\begin{split}
H_{\mu\nu}=2\bigg[RR_{\mu\nu}-2R_{\mu\omega}R^{\omega}_{\ \nu}-2R^{\omega\beta}R_{\mu\omega\nu\beta} & \\
+R_{\mu}^{\ \omega\beta\gamma}R_{\nu\omega\beta\gamma}\bigg]-\frac{1}{2}g_{\mu\nu}\mathcal{L}_{GB}.
\end{split}
\end{equation}
The energy-momentum tensor $T_{\mu\nu}$ corresponding to matter fields is obtained from $S_{matter}$. \\

So, by taking $n=5$, the $5$-dimensional line element for a static spherically symmetric spacetime has the standard form
\begin{eqnarray}\label{eq6}
\label{5} ds^{2}& =& -e^{2\nu(r)} dt^{2} + e^{2\lambda(r)} dr^{2} +
 r^{2}(d\theta^{2} + \sin^{2}{\theta} d\phi^2\nonumber\\
&& +\sin^{2}{\theta} \sin^{2}{\phi^2}d\psi),
\end{eqnarray}
in coordinates ($x^i = t,r,\theta,\phi,\psi$). For our model the energy-momentum tensor for the stellar fluid is taken to be
\begin{equation}\label{eq7}
T_{\mu\nu}=diag\left(-\rho,p_{r},p_{t},p_{t},p_{t}\right),
\end{equation}
where $\rho$, $p_{r}$, and $p_{t}$ are the proper energy density, radial pressure, and tangential pressure, respectively. By considering the comoving fluid velocity as $u^a=e^{-\nu}\delta_0^a$,
the EGB field equation (\ref{eq3}) leads to the following set of independent equations

\begin{figure}[t]
    \centering
        \includegraphics[scale=.6]{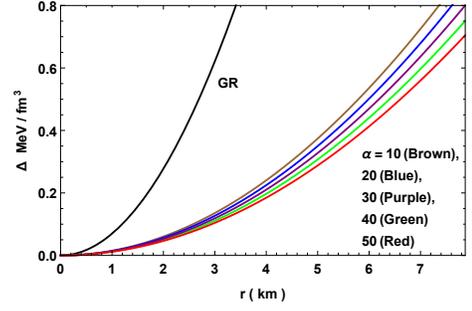}
       \caption{ Variation of anisotropy for 4U 1538-52 using the parameter provided in Table I.}
    \label{fig4}
\end{figure}

\begin{eqnarray}
\label{eq8}\kappa\rho &=& \frac{3}{e^{4\lambda} r^{3}} \left(4\alpha \lambda'+re^{2\lambda}-re^{4\lambda}- r^2 e^{2\lambda}\lambda' -4\alpha e^{2\lambda}\lambda'\right), \nonumber \\
\\
\label{eq9} \kappa p_r & = &\frac{3}{e^{4\lambda }r^3}
\left[-re^{4\lambda}+(r^2 \nu' +r +4\alpha \nu')e^{2\lambda} -4\alpha \nu'\right] , \\
\label{eq10} \kappa p_t &=& \frac{1}{e^{4\lambda }r^2} \left(- e^{4\lambda}- 4\alpha \nu''+ 12 \alpha \nu' \lambda' -4 \alpha(\nu')^2\right)
\nonumber\\
&& +\frac{1}{e^{2\lambda }r^2} \left(1- r^2 \nu' \lambda' +2r \nu'-2r  \lambda' +r^2(\nu')^2 \right) \nonumber \\
&& +\frac{1}{e^{2\lambda }r^2} \left(r^2 \nu'' -4\alpha
\nu'\lambda' + 4\alpha (\nu')^2 +4\alpha \nu''\right).
\end{eqnarray}
Besides, we have considered units such that the speed of light $c$ and the constant $G_{5}$ are set to unity. Then, $\kappa=8\pi$. Here, $\prime$ denotes  differentiation with respect to the radial coordinate $r$.

\section{Solution of the Field Equations}\label{sec3}
To solve the above field equations (\ref{eq8})-(\ref{eq10}) we choose,
$\lambda(r)= \ln(1 + ar^2 + br^4)$, and $\nu = Br^2 + 2 \ln C$ with $a$, $b$, $B$ and $C$ as constants. These metric potentials conform the well known Tolman-Kuchowicz \cite{tolman,kucho} spacetime. This election on $e^{\lambda}$ and $e^{\nu}$ is well motivated because both metric potentials are free from physical and mathematical singularities at every point inside the compact star. Moreover, at the center of the structure they have the appropriated behaviour \i.e $e^{\lambda(r)}|_{r=0}=1$ and $e^{\nu(r)}|_{r=0}=C^{2}$ which implies $(e^{\lambda})^{\prime}|_{r=0}=(e^{\nu})^{\prime}|_{r=0}=0$, as is required for a well behaved model. The trend of the inner geometry is displayed in the upper panel of Fig. \ref{fig1}. A completely regular behavior is observed, also as $\alpha$ grows $e^{\lambda}$ and $e^{\nu}$ take higher values, in distinction with GR whose values are dominated by those of EGB theory for all $r$. So, inserting $e^{\lambda}$ and $e^{\nu}$ into equations (\ref{eq8})-(\ref{eq10}) we arrive to

\begin{figure}[t]
    \centering
     \includegraphics[scale=.5]{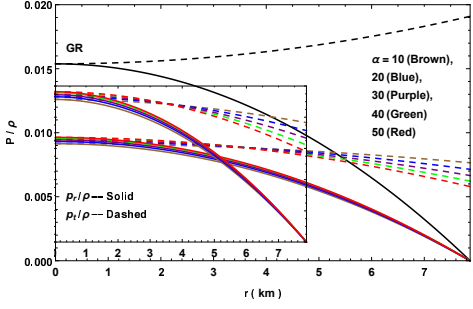}
       \caption{ Variation of equation of state parameters for 4U 1538-52 using the parameter provided in Table I.}
    \label{fig5}
\end{figure}

\begin{eqnarray} \label{eq11}
\kappa\rho &=& \frac{3}{r^3 \Psi^4}\Big[-\frac{8 \alpha r (a + 2 b r^2)}{\Psi} + 8 \alpha r (a + 2 b r^2) \Psi \nonumber\\
&& + 2 r^3 (a + 2 b r^2) \Psi - r \Psi^2 +  r \Psi^4)\Big]\\ \label{eq12}
    \kappa p_r &=& \frac{3}{r^3 \Psi^4}\Big[-8 \alpha B r + (r + 8 \alpha B r + 2 B r^3) \Psi^2 -  r \Psi^4\Big]\nonumber\\
    \\ \label{eq13}
   \kappa p_t &=& \frac{1}{\Psi^5}\Big[48 \alpha B (a + 2 b r^2) +8 \alpha B (a - 2 B + b r^2) \Psi \nonumber\\&&-4 \Big(a - 2 B + 2 a \alpha B + B_1 r^2\Big) \Psi^2 - (a + b r^2) \Psi^4 \nonumber\\&&- \Psi^3 \Big(a + 2 B + b r^2 - 4 B^2 (4 \alpha + r^2)\Big)\Big].
    \end{eqnarray}
The anisotropic factor defined by $\Delta\equiv p_{t}-p_{r}$ is obtained as,
\begin{eqnarray}\label{eq14}
\kappa\Delta &=& \frac{2}{\Psi^5} \Big[24 \alpha B (a + 2 b r^2) -8 \alpha B (a + B + b r^2) \Psi \nonumber\\&& -2 (a - 2 B + 8 a \alpha B + B_2 r^2) \Psi^2 + (a + b r^2) \Psi^4 \nonumber\\&&+ \Psi^3 \Big(a + b r^2 + 2 B (-2 + 4 \alpha B + B r^2)\Big)\Big]
\end{eqnarray}
 where,
\begin{eqnarray*}
B_1&=&-a B + b (2 + 6 \alpha B)\\
\Psi&=&(1 + a r^2 + b r^4)\\
B_2&=&-a B + 2 b (1 + 6 \alpha B)
\end{eqnarray*}

\begin{figure}[t]
    \centering
        \includegraphics[scale=.6]{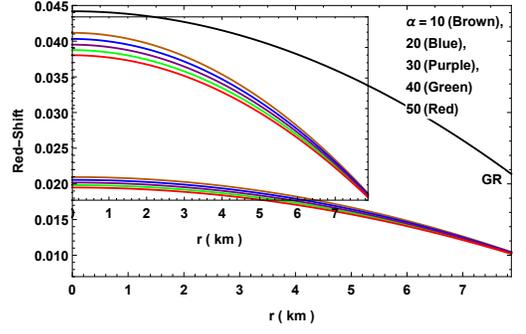}
       \caption{ Variation of gravitation red-shift for 4U 1538-52 using the parameter provided in Table I.}
    \label{fig6}
\end{figure}

The behavior of the metric function, density, pressure, anisotropy and equation of state parameter are given in Fig. \ref{fig1}-\ref{fig5}.
The interior red-shift can be found as
\begin{equation}
z(r) = e^{-\nu/2}-1
\end{equation}
and its trend in shown in Fig. \ref{fig6}.

\section{Physical Analysis}\label{sec4}

In this section we study and analyze the behaviour of the main physical salient features of the model. These correspond to the thermodynamic variables \i.e the energy-density $\rho$, the radial $p_{r}$ and tangential $p_{t}$ pressure . It is well known that a purely theoretical well behaved compact object from the physical and mathematical  point of view must satisfies some general requirements in order to contrast the astrophysical observational data. Mainly, these general criteria say that thermodynamic parameters must be monotonic decreasing functions at all points within the configuration from the center towards the surface of the object. Obviously, such behavior means that the maximum value of each of these physical quantities is attained at the center of the star. So, by means of second derivative criteria we have

\begin{eqnarray}
\kappa\frac{d\rho}{dr} &=& -\frac{6r}{\Psi^6}\Big[(a^2 + b +  a b r^2) \Psi^4 + 2 (a^2 + 8 b + a b r^2) \Psi^3 \nonumber\\
&&+40 \alpha (a^2 - 4 b) (a + b r^2)+8 a \alpha \Psi \Big\{4 a (a + b r^2)  \nonumber\\
&& +b\Big\}+ b \Psi^5+2 \Psi^2 \Big\{3 a^2 + 52 a \alpha b +12 b (4 \alpha b r^2 \nonumber\\
&& -1)\Big\}\Big] \\
\kappa\frac{dp_r}{dr} &=& \frac{6r}{\Psi^5}\Big[-32 \alpha B (a^2 - 2 b + a b r^2) +(a^2 + b + a b r^2) \nonumber\\
&&  \Psi^3 + 2 \Big\{A_3 + b (a - 4 B) r^2\Big\} \Psi^2 +b \Psi^4 - 8 \alpha B \Psi \nonumber\\
&& \Big\{b + 3 a (a + b r^2)\Big\}\Big]\\
\kappa\frac{dp_t}{dr} &=& \frac{2r}{\Psi^6}\Big[-240 \alpha (a^2 - 4 b) B -32 \alpha B \Big\{a^2 + 25 b -  \nonumber\\
&& 2 a B+ b (a - 4 B) r^2\Big\} \Psi + (a^2 + b - 12 B^2 + a b r^2) \nonumber\\
&& \Psi^4+b \Psi^5 +4 \Psi^2 \Big\{-6 a \alpha b B r^2 +3 a^2 (1 + 2 \alpha B  \nonumber\\
&& + B r^2)-2 b (6 + 25 \alpha B + 6 B r^2)\Big\} +  2 \Psi^3 \Big\{a^2 +  \nonumber\\
&& 18 b- 8 a B (1 + 2 \alpha B) +a (b + 4 B^2) r^2 \nonumber\\&&+ 4 B \big[2 B + b r^2 + \alpha b (15 - 8 B r^2)\big]\Big\}\Big].
\end{eqnarray}
Then, at the center of the star
 \begin{eqnarray*}
 \kappa\rho'' &=& -18 (1 + 8 a \alpha) (3 a^2 - 2 b)<0\\
 \kappa p_r'' &=& -6 \Big(4 a B + 2 b (1 - 8 \alpha B) + a^2 (-3 + 56 \alpha B)\Big)<0\\
 \kappa p_t'' &=& a^2 (30 - 496 \alpha B) + 32 a B (-1 + 2 \alpha B)\nonumber\\&& +
 4 \Big(2 B^2 + 5 b (-1 + 8 \alpha B)\Big)<0.
 \end{eqnarray*}
Additionally, in order to ensure a positive defined $\rho$, $p_{r}$ and $p_{t}$ throughout the compact object, the central density and central pressure must be positive at $r=0$. Hence, from Eqs. (\ref{eq8})-(\ref{eq10}) we get

\begin{figure}[t]
    \centering
        \includegraphics[scale=.6]{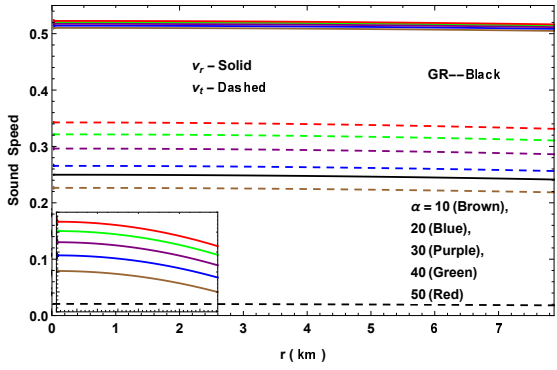}
        \caption{ Variation of sound speed for 4U 1538-52 using the parameter provided in Table I.}
    \label{fig7}
\end{figure}

\begin{eqnarray}\label{eq18}
\kappa\rho_c&=&12 a (1 + 4 a \alpha)>0\\ \label{eq19}
\kappa p_c&=&6 (B + a (-1 + 8 \alpha B))>0
\end{eqnarray}
It is clear from Eqs. (\ref{eq18})-(\ref{eq19}) that
\begin{equation}\label{eq20}
a>0 \quad \mbox{and} \quad B>\frac{a}{1+8a\alpha}.
\end{equation}
Moreover, it is also required to ensure that any physical fluid satisfies the Zeldovich's criterion i.e. $p_{rc}/ \rho_c \le 1$ which implies
\begin{equation}\label{eq21}
\frac{p_{rc}}{\rho_c} = \frac{B+a\left(8B\alpha-1\right)}{2a\left(1+4a\alpha\right)} \leq 1.
\end{equation}
On the other hand, regarding the anisotropy factor $\Delta=p_{t}-p_{r}$ given by Eq. (\ref{eq14}), plays an important role in the stellar matter distribution. In the study of anisotropic matter distributions, a well behaved model has a monotonic increasing anisotropy factor with increasing radial coordinate $r$ at all its interior points. It means that $\Delta>0$ everywhere within the star. Explicitly the former reads
\begin{equation}\label{eq22}
\Delta(r)= \left\{ \begin{array}{lcc}
             0, &   \mbox{if}  & r=0 \\
\\ p_{t}(R), &  \mbox{if} & r=R.
\end{array}\right.
\end{equation}

The first statement of (\ref{eq22}) ensures the regularity of the solution in the origin, besides that this allows the possibility that the inner geometry is regular not only at the center of the structure but in all points. Vanishing $\Delta$ at the center purports $p_{r}(0)=p_{t}(0)$. This fact is a consequence of matter collineation induced by the Killing vector fields of the spherical symmetry. The second statement of (\ref{eq22}) says that at the boundary of the star $\Sigma$ (defined by $r=R$, where $R$ stand for the radius of the object), $p_{r}(R)=0$ and in consequence $\Delta(R)=p_{t}(R)>0$ implying $\Delta(r)>0$ for $0\leq r\leq R$. As  pointed out early one needs positive thermodynamic variables throughout the star. Furthermore, a positive anisotropy factor introduces in the system a repulsive force (outward) that helps to counteract the gravitational gradient. The presence of a repulsive anisotropic force allows the construction of more compact objects \cite{r41}. In addition, it contributes to enhance the equilibrium and stability mechanism. From Figs. \ref{fig2} and \ref{fig3} for different values of the parameter $\alpha$,  we can see the behaviour of the energy density $\rho$ (Fig. \ref{fig2}), both the radial $p_{r}$ and transverse $p_{t}$ pressure (Fig. \ref{fig3}) and the anisotropy factor $\Delta$ (Fig. \ref{fig4}) against the radial coordinate inside the star. It is observed that for $10\leq \alpha \leq 50$ (EGB gravity) the maximum values reached by all the physical quantities is less than the values reached by GR theory ($\alpha\rightarrow 0$). Moreover, the anisotropy factor is greater in GR theory than in EGB theory. We also observe that as $\alpha$ increases the anisotropy decreases at each interior point of the configuration. The effect of the EGB term is to diminish the relative difference between the radial and tangential stresses. This may be a possible mechanism to achieve pressure isotropy within the stellar interior. Besides, Fig. \ref{fig5} shows the behaviour of the ratios $p_{r}/\rho$ and $p_{t}/\rho$. It is appreciated that Zeldovich's condition is satisfied.

\begin{figure}[t]
    \centering
        \includegraphics[scale=.6]{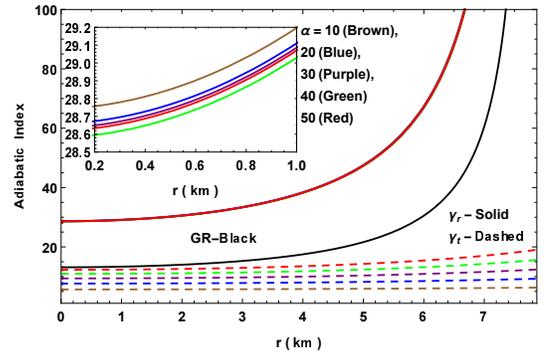}
       \caption{ Variation of adiabatic index for 4U 1538-52 using the parameter provided in Table I.}
    \label{fig8}
\end{figure}

\section{Exterior spacetime and matching conditions}\label{sec5}

In order to obtain the constant parameters that characterize the model \i.e $a$, $b$, $B$ and $C$ it is necessary to match in a smoothly way the internal manifold $\mathcal{M}^{-}$ given by Tolman-Kuchowicz \cite{tolman,kucho} spacetime with the static exterior spacetime $\mathcal{M}^{+}$ in $5$-D which is described by the Einstein-Gauss-Bonnet-Schwarzschild solution \cite{r4}
\begin{eqnarray}
\label{eq23} ds^{2}& =& -F(r) dt^{2} + [F(r)]^{-1} dr^{2} +
 r^{2}\left(d\theta^{2} + \sin^{2}{\theta} d\phi^2\right.\nonumber\\
 &&\left. +\sin^{2}{\theta} \sin^{2}{\phi}
 d\psi\right),
\end{eqnarray}
where,
\begin{equation}
\label{eq24} F(r) =1+\frac{r^2}{4 \alpha}
\left(1-\sqrt{1+\frac{8\alpha M}{r^4}}\right).
\end{equation}

\begin{figure}[t]
    \centering
         \includegraphics[scale=.6]{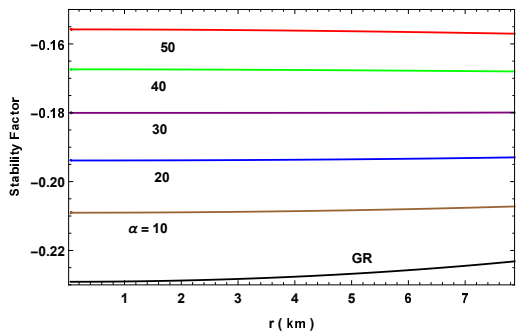}
       \caption{ Variation of stability factor for 4U 1538-52 using the parameter provided in Table I.}
    \label{fig9}
\end{figure}

In (\ref{eq24}) $M$ is associated with the gravitational mass of the hypersphere. It is remarkable to mention that when $\alpha\rightarrow 0$ the usual Schwarzschild solution is recovered. So, joining the inner and exterior spacetime demands the compliance of the first and second fundamental forms. These matching conditions are known as Isarel-Darmois  junction conditions \cite{r44,r45}. The first fundamental from consists in the continuity of the metric potential and its derivative across the boundary $\Sigma$. Explicitly
\begin{equation}\label{eq25}
\left[ds^{2}\right]_{\Sigma}=0,
\end{equation}
\begin{equation}\label{eq26}
e^{\lambda^{-}}|_{r=R}=e^{\lambda^{+}}|_{r=R} \quad \mbox{and} \quad   e^{\nu^{-}}|_{r=R}=e^{\nu^{+}}|_{r=R},
\end{equation}
and
\begin{equation}\label{eq27}
\left(\frac{\partial e^{\nu^{-}}}{\partial r}\right)_{|r=R}=\left(\frac{\partial e^{\nu^{+}}}{\partial r}\right)_{|r=R}.
\end{equation}
The second fundamental form reads
\begin{equation}\label{eq28}
p_{r}(R)=0.
\end{equation}
This condition determines the object size. This is so because, the pressure decreases as we approach to the surface and the pressure at the exterior of the star must be null, then this will correspond to the star boundary. In other words second fundamental form says that the matter distribution is confined in a finite spacetime region, in consequence the star does not expand indefinitely beyond $\Sigma$. Therefore, from first fundamental form we obtain
\begin{eqnarray}\label{eq29}
\frac{1}{1+aR^2+bR^4}&=&1+\frac{R^2}{4 \alpha}
\left(1-\sqrt{1+\frac{8\alpha M}{R^4}}\right),\\ \label{eq30}
C^2e^{BR^{2}}&=&1+\frac{R^2}{4 \alpha}
\left(1-\sqrt{1+\frac{8\alpha M}{R^4}}\right),\\ \label{eq31}
2B C^2e^{BR^{2}}&=&-\frac{1}{2\alpha}\frac{1-\sqrt{1+\frac{8\alpha M}{R^{4}}}}{\sqrt{1+\frac{8\alpha M}{R^{4}}}},
\end{eqnarray}
and from the second fundamental form we get

\begin{figure}[t]
    \centering
        \includegraphics[scale=.6]{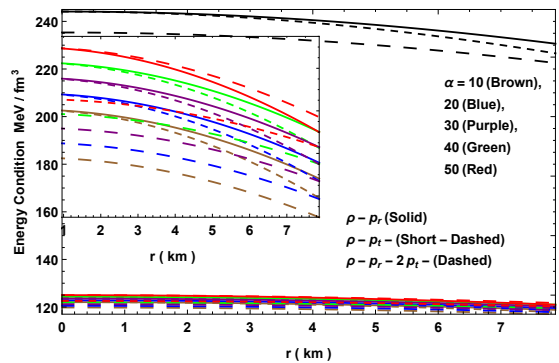}
       \caption{ Variation of energy conditions for 4U 1538-52 using the parameter provided in Table I.}
    \label{fig10}
\end{figure}

\begin{equation}\label{eq32}
\begin{split}
-8BR\alpha+\left(R+8BR\alpha+2BR^{3}\right)\left(1+aR^{2}+bR^{4}\right)^{2} &\\
-R\left(1+aR^{2}+bR^{2}\right)^{4}=0.
\end{split}
\end{equation}
Eqs. (\ref{eq29})-(\ref{eq32}) are the necessary and sufficient conditions to determine the complete set of constant parameters $\{a,b,B,C\}$ that describe the model.

\section{Causality Condition}\label{sec6}
For any model describing a stellar interior, the subliminal sound speed of the pressure waves must be less than the speed of light. In the treatment of anisotropic fluids, the propagation of the pressure waves is along the main directions of the object, \i.e the radial and transverse direction. The subliminal sound speeds along these directions are defined by
\begin{equation}
v_{r}=\sqrt{\frac{d{p}_{r}}{d{\rho}}} \quad \mbox{and} \quad v_{t}=\sqrt{\frac{d{p}_{t}}{d{\rho}}}.
\end{equation}
So, in order to obtain a physically admissible model, both speeds $v_{r}$ and $v_{t}$ must be bounded  by the speed of light. This is the so called causality condition. Causality means that pressure (sound) waves in the fluid do not propagate at arbitrary speeds. On the other hand, in distinction with what happens in the case of isotropic fluid (in this case the pressure waves propagate only in one direction because $p_{r}=p_{t}$), the speed behaviour within the star against the radial coordinate in decreasing. However, this is not true in the case where there is anisotropy, since the  speed behaviour depends on the rigidity of the material. So, causality condition reads
\begin{equation}\label{eq34}
0\leq v_{r}\leq 1 \quad \mbox{and}\quad  0\leq v_{t}\leq 1,
\end{equation}
where the speed of light was taken to be $c=1$. Preservation/no-preservation of causality condition (\ref{eq34}) has strong implications on the matter distribution within the structure. Because, it is related with the behaviour of the energy-momentum tensor that describes the material content. Preservation of causality yields a well defined energy-momentum tensor. Additionally, the fact of having different speeds in the directions mentioned above, influences the stability of the system.\\

\begin{figure}[t]
    \centering
        \includegraphics[scale=.5]{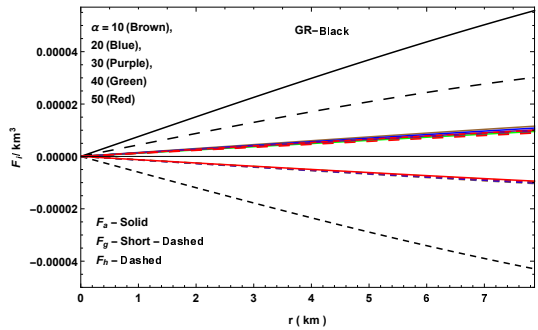}
       \caption{ Variation of various forces in TOV-equation for 4U 1538-52 using the parameter provided in Table I.}
    \label{fig11}
\end{figure}

Now at the center of the star the subliminal sound speeds are,
\begin{eqnarray}\label{eq35}
v_{r0}^2&=&\frac{4 a B + 2 b (1 - 8 \alpha B) +
 a^2 (-3 + 56 \alpha B)}{3 (1 + 8 a \alpha) (3 a^2 - 2 b)}\\ \label{eq36}
v_{t0}^2&=&\frac{1}{9 (1 + 8 a \alpha) (3 a^2 - 2 b)}\Big[-4 B^2 + b (10 - 80 \alpha B) \nonumber\\&&+ 16 a B (1 - 2 \alpha B) +
 a^2 (-15 + 248 \alpha B)\Big].
 \end{eqnarray}
Expressions (\ref{eq35})-(\ref{eq36}) impose some restrictions on $\{a,b,B\}$ in order to preserve  causality condition. Fig. \ref{fig7} (upper panel) exhibits the trend of both speeds throughout the configuration. As we can see EGB theory dominates GR in both directions. Besides, the speed of the radial and tangential subliminal sound pressure waves are decreasing in nature and as $\alpha$ increases they take higher values at the center of the star. Otherwise with increasing radius.

\section{Stability mechanisms: relativistic adiabatic index and Abreu's criterion}\label{sec7}

In this section we analyze an important mechanism. We refer to stability mechanism. The general theory of stability is made complicated since many variables can change at the same time. Therefore maintaining consistency can be a difficult task. Within this branch there are some heuristic methods of determining stability, such as relativistic adiabatic index \cite{bondi,heintz}, Abreu's criterion \cite{r46} (based on Herrera's cracking concept \cite{r35}), static stability criterion \cite{harri,zel}, Ponce De Leon's criterion \cite{ponce}, among others. This clearly suggests that the study of stability of compact objects can only be carried out in a timely manner, that is, there is no mechanism to test whether an astrophysical system is stable from a global point of view. However, these heuristic mechanisms, even if only in a timely manner, allow us to check how stable an anisotropic matter distribution is, which is susceptible to radial disturbances due to the presence of repulsive forces in case $\Delta>0$ (in the case of attractive forces, which occurs when $\Delta<0$ the system is also under disturbances of the radial type.). So, in this opportunity we use the first two \i.e the relativistic adiabatic index and the Abreu's criterion. The former gives the ratio of two specific heats, it is defined
by

\begin{figure}[t]
    \centering
        \includegraphics[scale=.55]{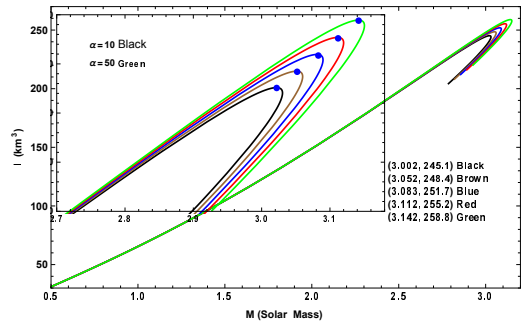}
       \caption{The $M-I$ graphs is plotted for $a=2.13\times 10^{-4}$ and $b=4\times 10^{-8}$.}
    \label{fig13}
\end{figure}

\begin{figure}[t]
    \centering
        \includegraphics[scale=.5]{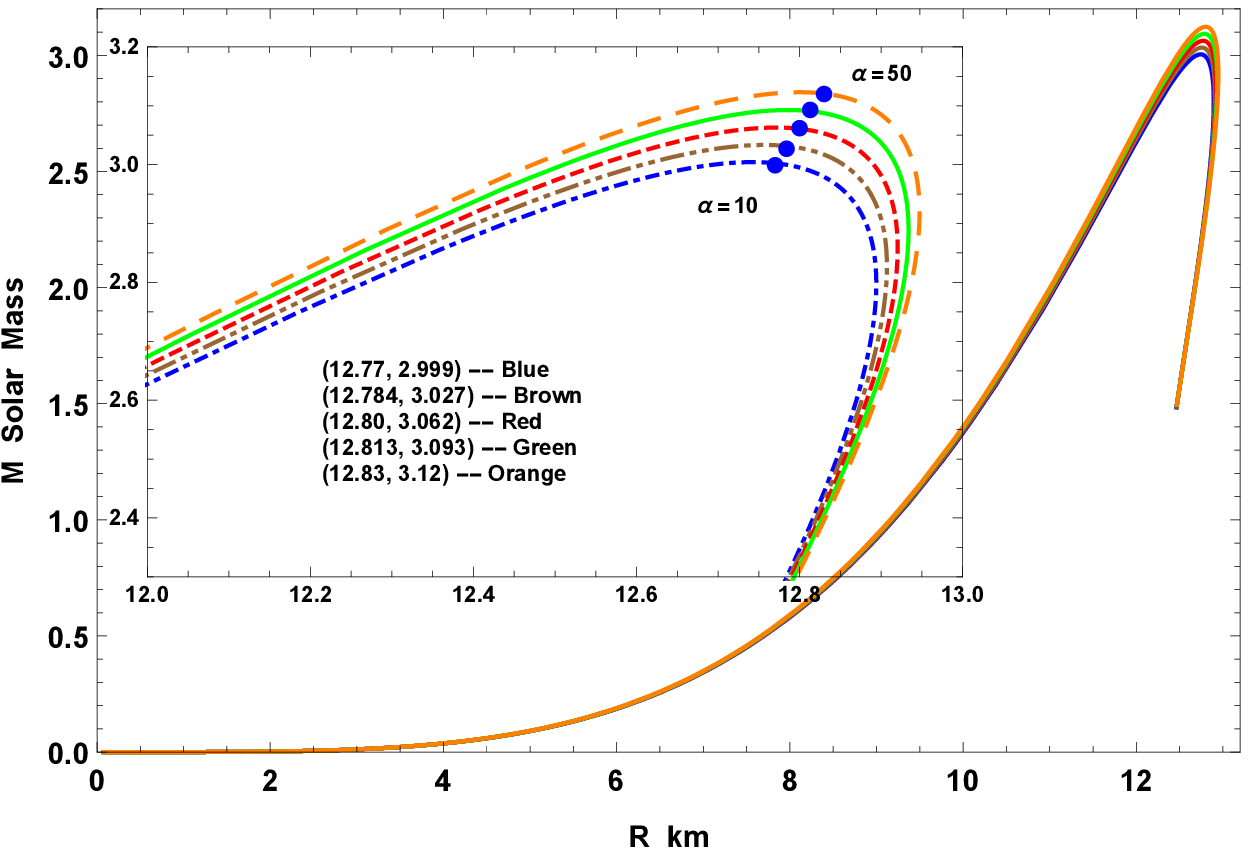}
       \caption{The $M-R$ graphs is plotted for $a=2.13\times 10^{-4}$ and $b=4\times 10^{-8}$.}
    \label{fig12}
\end{figure}

\begin{eqnarray}\label{eq37}
\Gamma_r &=& \frac{\rho+p_r}{p_r}v_r^2 \nonumber\\
&=& \frac{\chi_1 \Big\{1 + (4 \alpha + r^2) (a + b r^2) (2 + a r^2 +b r^4)\Big\}}{\Psi \big[2 B - (a + b r^2) (2 + a r^2 + b r^4) \chi_2\big]}v_r^2\nonumber\\
\\ \label{eq38}
 \Gamma_t &=& \frac{\rho+p_t}{p_t}v_t^2 \nonumber\\
&=&\frac{2}{\zeta}\Big[A_1 + 3 A_2 r^2 +2 A_3 r^4 + A_4 r^6 + A_5 r^8 +A_6 r^{10} \nonumber\\
&& +2 b^2 A_7 r^{12} + 4 a b^3 r^{14} + b^4 r^{16} - \chi_3\Big]v_t^2,
\end{eqnarray}
 being,
 \begin{eqnarray*}
\chi_1 &=& 2 \Big(a + B + (2 b + a B) r^2 +b B r^4\Big)\\
 \chi_2 &=& 2 a r^2- 8 \alpha B - 2 B r^2 + r^4 (2 b + [a + b r^2]^2)  +1 \\
\chi_3 &=& \frac{12 \alpha \Big(a + 4 B + 2 (b + a B) r^2\Big)}{1 + a r^2 + b r^4}\\
\chi_4 &=& \frac{48 \alpha B r^2 (a + 2 b r^2)}{1 + a r^2 + b r^4}\\
\chi_5 &=& \frac{-3 + 2 B r^2 + (4 + 2 a r^2)}{
1 + a r^2 + b r^4}\\
\chi_6 &=&-7 b r^2 +
4 B \Big(1 - b r^4 + B (r^2 + b r^6)\Big)\\&&+ a (-3 + 4 B^2 r^4) \\
\zeta &=&  \chi_4 -8 \alpha B - 16 \alpha B^2 r^2 - \Psi^4  \\
&& +2 B \Psi^2 (r^2 +4 \alpha \chi_5)+ \Psi(1 +r^2 \chi_6)
\end{eqnarray*}

where the constants $A_i's$ ($i=1,2,...,7$) are given by,
\begin{eqnarray*}
A_1 &=& 12 \alpha (a + 4 B)\\
A_2 &=& a + 4 a^2 \alpha + 8 \alpha b + B\\
A_3 &=& B^2 + b (2 - 4 \alpha B) + a^2 (3 - 2 \alpha B) + 2 a (9 \alpha b  \nonumber \\
&& + B+ 4 \alpha B^2)\\
A_4 &=& 4 a^3 + 13 a b + 24 \alpha b^2 + (a^2 + 2 b - 16 a \alpha b) B \\
&& + 4 (a + 2 a^2 \alpha + 4 \alpha b) B^2\\
A_5 &=& 12 a^2 b +7 b^2 - 12 \alpha b^2 B +
2 (a^2 + 2 b + 8 a \alpha b) B^2 \nonumber \\
&& + a^4\\
A_6 &=& b \Big[4 a^3 + b B (-1 + 8 \alpha B) + 4 a (3 b + B^2)\Big]\\
A_7 &=& 3 a^2 + 2 b + B^2
\end{eqnarray*}
The expressions (\ref{eq37})-(\ref{eq38}) provide the relativistic adiabatic index in the radial $\Gamma_{r}$ and tangential $\Gamma_{t}$ direction. However, it should be noted that in the event of a gravitational re-collapse of the structure, it is sufficient to study the behavior of the relativistic adiabatic index in the radial direction, since the compression of the object due to the gravitational force would occur in that direction. Bondi's pioneering work \cite{bondi} has shown that $\Gamma>4/3$ is the condition for the stability of a Newtonian isotropic matter distribution. This condition is very different in the case of anisotropic relativistic fluid spheres, because the stability will depend on the type of anisotropy. Then, the stability condition for  an anisotropic relativistic sphere,
is given by \cite{r32,r33,r34}
\begin{equation}\label{eq39}
\Gamma>\frac{4}{3}+\left[\frac{1}{3}\kappa\frac{\rho_{0}p_{r0}}{|p^{\prime}_{r0}|}r+\frac{4}{3}\frac{\left(p_{t0}-p_{r0}\right)}{|p^{\prime}_{r0}|r}\right]_{max}
\end{equation}
where $\rho_{0}$, $p_{r0}$ and $p_{t0}$ are the initial density, radial and tangential pressure when the fluid is in static equilibrium. The second term in the right hand side represents the relativistic corrections to the Newtonian perfect fluid and the third term is the contribution due to anisotropy. It is clear from (\ref{eq39}) that if we have a non-relativistic perfect fluid matter distribution the bracket vanishes and we recast the collapsing Newtonian limit $\Gamma<4/3$. Heintzmann and Hillebrandt \cite{heintz} showed that in the presence of a positive an increasing anisotropy factor $\Delta=p_{t}-p_{r}>0$, the stability condition for a relativistic compact object is given by $\Gamma>4/3$, that is so because positive anisotropy factor may slow down the growth of instability. In Fig. \ref{fig8} it has been shown that $\Gamma_{r}, \Gamma_{t}>4/3$ everywhere within the stellar interior for both EGB and GR theories. Therefore, from the  relativistic adiabatic index point of view our model is stable. \\

On the other hand, Abreu's criterion \cite{r46}, basically consists in contrasting the speed of pressure waves in the two principal directions of the spherically symmetric star: the subliminal radial sound speed with the subliminal tangential sound speed, and then based on those values at particular points in the object, one could potentially conclude whether the system is stable or unstable under cracking instability. Cracking process is mechanism to study  instability when anisotropy matter distributions are present \cite{r35}. Nevertheless, this mechanism can be characterized most easily through the subliminal speed of pressure waves. Furthermore, from causality condition one has $0\leq v^{2}_{r}\leq 1$ and $0\leq v^{2}_{t}\leq 1$ which implies $0\leq |v^{2}_{t}-v^{2}_{r}|\leq 1$. Explicitly it reads
\begin{eqnarray}
     \label{eq41}
   & \quad\hspace{-6.2cm} -1\leq v^{2}_{t}-v^{2}_{r}\leq 1  =\nonumber\\ &
\quad\hspace{0.2cm} \left\{
	       \begin{array}{ll}
		   -1\leq v^{2}_{t}-v^{2}_{r}\leq 0~~ & \mathrm{Potentially\ stable\ }  \\
		 0< v^{2}_{t}-v^{2}_{r}\leq 1 ~~ & \mathrm{Potentially\ unstable}
	       \end{array}
	        \right\}.~~
	    \end{eqnarray}
So, the principal aim of Abreu's criterion is that if the subliminal tangential speed $v^{2}_{t}$ is larger than the subliminal radial speed $v^{2}_{r}$; then instabilities regions may occur in the object, rendering the latter an unstable configuration. So, with the help of graphical analysis one can determines the potentially stable/unstable regions inside the star and then conclude whether the system is stable or not, at least locally. From Fig. \ref{fig9} it is appreciated that the system presents all the regions completely stable for all values of $\alpha$, including GR. However, GR seems to be more stable than EGB theory, because the stability factor takes smaller values between $-1$ and $0$ respect to EGB theory.

\begin{table*}
\caption{\label{tab} Values of all the parameters corresponding to different values of $\alpha$ and the corresponding $(M_{max},R)$.}
\centering
\begin{ruledtabular}
\begin{tabular}{llllllllllp{0.55in}}
$\alpha$ & $b$ & $a$ & $B$ & $c$ & $\rho_c$ & $p_c$ & $\rho_b$ & $M_{max}$ & $R$ \\
& $\times 10^{-8} (km^{-4})$ & $\times 10^{-3}(km^{-2})$ &$\times 10^{-3}~(km^{-2})$ & & $\times 10^{14}~(gm.cm^{-3})$ & $\times 10^{33}~(dyne.cm^{-2})$ & $\times 10^{14}~(gm.cm^{-3})$ & $(M_\odot)$ & (km)\\
\hline
0  & 15& 688655 & 713032 & 0.95768  & 4.43142 & 7.04898 & 4.09734 &- & -\\
10 & 4 & 33681  & 33405  & 0.97947  & 2.19653 & 1.80466 & 2.11426 & 2.999 & 12.77\\
20 & 4 & 334567 & 32364  & 0.979853 & 2.21053 & 1.84999 & 2.13184 & 3.027 & 12.784 \\
30 & 4 & 332382 & 3047   & 0.980493 & 2.22415 & 1.88714 & 2.14398 & 3.062 & 12.80\\
40 & 4 & 330251 & 31387  & 0.980278 & 2.23742 & 1.92088 & 2.15641 & 3.093 & 12.813\\
50 & 4 & 328172 & 29605  & 0.980881 & 2.25035 & 1.95076 & 2.16905 & 3.12 & 12.83\\
\end{tabular}
\end{ruledtabular}
\end{table*}

\section{energy conditions}\label{sec8}
The matter content that makes up astrophysical bodies can be composed of a large number of material fields. Although the components that constitute the matter distribution are known, it could be very complex to describe the concrete form of the energy-momentum tensor. Indeed, one has some ideas on the behaviour of the matter under extreme conditions of density and pressure. \\

Nonetheless there are certain inequalities which are physically reasonable to assume to check the conduct of the energy-momentum tensor at every point inside the star. These inequalities are known as energy conditions. Then we have the null energy condition (NEC), strong energy condition (SEC) and weak energy condition(WEC). Explicitly, these are given by
\begin{eqnarray}
\text{WEC} &:& T_{\mu \nu}l^\mu l^\nu \ge 0~\mbox{or}~\rho \geq  0,~\rho+p_i \ge 0 \\
\text{NEC} &:& T_{\mu \nu}t^\mu t^\nu \ge 0~\mbox{or}~ \rho+p_i \geq  0\\
\text{SEC} &:& T_{\mu \nu}l^\mu l^\nu + {1 \over 2} T^\lambda_\lambda l^\sigma l_\sigma \ge 0 ~\mbox{or}~ \rho+\sum_i p_i \ge 0
\end{eqnarray}
where $i\equiv (radial~r, transverse ~t),~l^\mu$ and $t^\mu$ are time-like vector and null vector respectively. To verify a well behaved energy-momentum tensor everywhere within the compact structure the above inequalities must satisfy simultaneously. In Fig. \ref{fig10}, we have plotted the L.H.S of the above inequalities which verifies that  all the energy conditions are satisfied at the stellar interior.\\

Moreover, from the physical point of view NEC means that an observer traversing a null curve will measure the ambient (ordinary) energy density to be positive. WEC implies that the energy density  measured by an observer crossing a timelike curve is never negative. SEC purports that the trace of the tidal tensor measured by the corresponding observers is always non-negative \cite{r47}. Furthermore, violations of energy conditions have sometimes been presented as only being produced by unphysical stress energy tensors. Usually SEC is used as a fundamental guide will be extremely idealistic. Nevertheless, SEC is violated in many cases, e.g. minimally coupled scalar field and curvature-coupled scalar field theories. It may or may not imply the violation of the more basic energy conditions i.e. NEC and WEC.

\section{generalized Tolman-Oppenheimer-volkoff equation}\label{sec9}

In this section we discuss the dynamical equilibrium condition of the stellar model by using Tolman-Oppenheimer-Volkoff  (TOV) in five dimensions \cite{tolman,r48} equation, given by
  \begin{eqnarray}\label{eq44}
 -\frac{dp_r}{dr}-\frac{\nu^{{\prime}}}{2} \big( \rho + p_{r} \big)+\frac{3}{r}\left({p_t}-{p_r}\right)=0,
\end{eqnarray}
where we denote first term $-\frac{dp_r}{dr}=F_h$, second term $-\frac{\nu^{{\prime}}}{2} \big( \rho+p_r \big)=F_g$ and third term $\frac{3}{r}\left({p_t}-{p_r}\right)=F_{a}$. These terms describe the hydrostatic force ($F_h$), gravitational force ($F_g$) and anisotropic force ($F_a$), respectively. \\

In the case of isotropic fluid spheres (${p}_{r}={p}_{t}$) and regarding $\alpha\rightarrow 0$ (GR limit), this equation drives the equilibrium of relativistic compact structures described by isotropic matter distribution. Regarding the presence of anisotropies and the EGB framework, this equation still drives the balance of the system. As was pointed out before, the present model is under three forces.  To guarantee the equilibrium of the proposed stellar structure, we have shown in Fig. \ref{fig11} that the balance of the forces is reached to all the values of ${\alpha}$ and GR also. Consequently, Fig. \ref{fig11} indicates that, in the situation of $10\leq\alpha\leq 50$, the resulting impact of hydrodynamic force (${F_h}$) and anisotropic force (${F_a}$)  compensates the internal attraction due to the gravitational force (${F_g}$). Furthermore, it is worth mentioning that in the GR case the $F_{h}$, $F_{a}$ and $F_{g}$ forces are greater than the corresponding EGB forces.

\section{Rigid Rotation, Moment of inertia and comparison with $M-R$ Graph}

Bejger and Haensel \cite{bejg} proposed  an approximate formula which convert a static model to rotating model and is given below:
\begin{equation}
I = {2 \over 5} \Big[1+{(M/R) \cdot km \over M_\odot}\Big] {MR^2}.
\end{equation}
Using the above expression we have plotted the trend of $I$ w.r.t. mass $M$ in Fig. \ref{fig13}. From this graph it can be seen that the maximum moment of inertia ($I_{max}$) increases with increase in coupling constant $\alpha$. Also from the $M-R$ graph (Fig. \ref{fig12}) we can also see that as $\alpha$ increases the maximum mass ($M_{max}$) also increases. From the two Figs. \ref{fig13} and \ref{fig12} one can notice that the sensitivity of $M-I$ graph is better than $M-R$ graph when stiffness of the equation of state changes.

\section{concluding remarks}\label{sec10}

It is evident that within the framework of 5-dimensional Einstein-Gauss-Bonnet gravity theory, it is plausible to obtain models that describe real compact objects such as white dwarfs, neutron stars and others. In addition, the obtained solution fulfils the basic and general requirements to be a physically and mathematically admissible model. In this opportunity we have solved the field equations (\ref{eq8})-(\ref{eq10}) by imposing the Tolman-Kuchowicz spacetime (Fig. \ref{fig1}). This election is well motivated for two reasons: $i)$ this metric is free from physical and geometrical singularities, so is completely plausible to describe the inner geometry of compact objects, $ii)$ yields to a well behaved energy density , \i.e positive defined and monotone decreasing function from the center to the boundary of the star (Fig. \ref{fig2}). As it is well known, this is a fundamental requirement to describe in a good way the material content inside the star.\\

Moreover, the remaining  thermodynamic variables that characterize the solution \i.e, the radial $p_{r}$ and tangential $p_{t}$ pressure are well behaved at all points within the configuration (\ref{fig3}). Besides, the tangential pressure $p_{t}$ coincides with the radial pressure $p_{r}$ at the center and then is always greater than $p_{r}$ everywhere. Actually, it is a very important fact, because it induces a positive anisotropy factor $\Delta$ inside the star (Fig. \ref{fig4}). A positive $\Delta$ brings with it important consequences for the structure. For example, it allows the construction of more compact objects (greater amount of mass contained in a smaller size) and introduces a force (repulsive in nature) that helps sustain the hydrostatic balance by counteracting the gravitational compression. The latter not only prevents the system from being subject to a gravitational re-collapse (as it would be in the case of an $\Delta<0$, which would introduce an attractive force, contributing to the gravitational gradient to collapse the object, which can take it even below its Schwarzschild's radius to form a black hole), but it improves the stability of the system as well.\\

On the other hand, as the material content is confined within the region given by $\Sigma=r=R$, to find all the constant parameters that describe the solution $\{a,b,B,C\}$ we have made the junction between the internal geometry and the outer spacetime, the Schwarzschild equivalent solution  (free of material content, \i.e vacuum solution) in EGB. This was performed by applying the first and second fundamental forms.\\

The remaining main physical highlights of the current solution can be summarized as follows:

\begin{enumerate}
\item Causality condition (Fig. \ref{fig7}) for the stability of the anisotropic matter distribution as a profile of the difference in squared of subliminal sound speed of the pressure waves, $| v^2_{t} - v^2_{r}|$ with respect to the radial coordinate $r$ satisfies the inequality $-1< v^2_{t} - v^2_{r}<0$ which manifests itself in Fig. \ref{fig9} (lower panel).

\item In Fig.\ref{fig8} we have displayed the behaviour of the adiabatic index~$\Gamma$ with respect to the infinitesimal radial adiabatic perturbation which confirms that when $\Gamma> 4/3$ our stellar structure is stable in all interior points of the stellar object with spherical symmetry.

\item Examination of the energy
conditions in order to test the physical validity of the obtained solution. In Fig. \ref{fig10} we have indicated the behaviour of all energy conditions with respect to the radial coordinate $r$ for the stellar system, which shows that our compact stellar structure is well satisfied for the system in the context of the EGB gravity  at various choose values of $\alpha$,  also considering GR theory.

\item We have shown in Fig. \ref{fig11} that the equilibrium of the forces is reached to all the values of $\alpha$ (including GR), which confirms that our stellar model is stable with respect to the equilibrium of forces.

\item
The stiffness of the corresponding EoS increases with increase in coupling constant $\alpha$, however, less stiff w.r.t. GR limit. The maximum mass corresponding to $\alpha=10$ to 50 is given in Table \ref{tab}. As $\alpha$ increases to 10 to 50, the moment of inertia also increases . This makes the EoS stiffer and therefore can support more  masses (Figs. \ref{fig13}, \ref{fig12}).

\end{enumerate}

Finally, it is worth mentioning that taking $\alpha\rightarrow 0$ GR results in 5-dimension are recovered. Moreover, as we can observe in the complete graphic analysis GR provides a more compact and stable model in distinction with EBG. Nevertheless, the same can be reached in the arena of EGB gravity taking smaller values of the coupling constant $\alpha$.

\begin{acknowledgments}
P.B is thankful to IUCAA, Govt of India for providing visiting associateship, F. Tello-Ortiz thanks the financial support by the CONICYT PFCHA/DOCTORADO-NACIONAL/2019-21190856, grant Fondecyt No. 1161192, Chile and project ANT-1855 at the Universidad de Antofagasta, Chile.
\end{acknowledgments}

\end{document}